\documentclass[twocolumn]{aastex63}
\usepackage{graphicx}	% Including figure files
\usepackage{amsmath}	% Advanced maths commands
\usepackage{amssymb}
\usepackage{verbatim}

\def\chandra{\textit {Chandra}}
\def\civ{C~{\sc iv}}
\def\aox{$\alpha_{\rm ox}$}
\def\daox{$\Delta\alpha_{\rm ox}$}
\def\lopt{$L_{\rm 2500~\AA}$}
\def\fopt{$f_{\rm 2500~\AA}$}
\def\fx{$f_{\rm 2~keV}$}

\def\lbol{$L_{\rm bol}$}
\def\xray{\hbox{X-ray}}

\received{}
\revised{}
\accepted{}
\submitjournal{ApJL}

\shorttitle{Extreme X-ray variability}
\shortauthors{Ni et al.}

\begin{document}

\title{An Extreme X-ray Variability Event of a Weak-Line Quasar}

\correspondingauthor{Qingling Ni}
\email{qxn1@psu.edu}

\author{Q. Ni}
\affiliation{Department of Astronomy \& Astrophysics, The Pennsylvania State University, 525 Davey Lab, University Park, PA 16802, USA}
\affiliation{Institute for Gravitation and the Cosmos, The Pennsylvania State University, University Park, PA 16802, USA}

\author{W. N. Brandt}
\affiliation{Department of Astronomy \& Astrophysics, The Pennsylvania State University, 525 Davey Lab, University Park, PA 16802, USA}
\affiliation{Institute for Gravitation and the Cosmos, The Pennsylvania State University, University Park, PA 16802, USA}
\affiliation{Department of Physics, The Pennsylvania State University, University Park, PA 16802, USA}

\author{W. Yi}
\affiliation{Department of Astronomy \& Astrophysics, The Pennsylvania State University, 525 Davey Lab, University Park, PA 16802, USA}
\affiliation{Yunnan Observatories, Kunming, 650216, China}
\affiliation{Key Laboratory for the Structure and Evolution of Celestial Objects, Chinese Academy of Sciences, Kunming 650216, China}

\author{B. Luo}
\affiliation{School of Astronomy and Space Science, Nanjing University, Nanjing, Jiangsu 210093, China}
\affiliation{Key Laboratory of Modern Astronomy and Astrophysics (Nanjing University), Ministry of Education, Nanjing, Jiangsu 210093, China}
\affiliation{Collaborative Innovation Center of Modern Astronomy and Space Exploration, Nanjing 210093, China}

\author{J. D. Timlin III}
\affiliation{Department of Astronomy \& Astrophysics, The Pennsylvania State University, 525 Davey Lab, University Park, PA 16802, USA}
\affiliation{Institute for Gravitation and the Cosmos, The Pennsylvania State University, University Park, PA 16802, USA}

\author{P. B. Hall}
\affiliation{Department of Physics \& Astronomy, York University, 4700 Keele Street, Toronto, ON M3J 1P3, Canada}

\author{Hezhen Liu}
\affiliation{School of Astronomy and Space Science, Nanjing University, Nanjing, Jiangsu 210093, China}
\affiliation{Key Laboratory of Modern Astronomy and Astrophysics (Nanjing University), Ministry of Education, Nanjing, Jiangsu 210093, China}
\affiliation{Collaborative Innovation Center of Modern Astronomy and Space Exploration, Nanjing 210093, China}

\author{R. M. Plotkin}
\affiliation{Department of Physics, University of Nevada, 1664 N. Virginia St, Reno, Nevada, 89557, USA}

\author{O. Shemmer}
\affiliation{Department of Physics, University of North Texas, Denton, TX 76203, USA}

\author{F. Vito}
\affiliation{Instituto de Astrof\'{i}sica and Centro de Astroingenier\'{i}a, Facultad de F\'{i}sica, Pontificia Universidad Cat\'{o}lica de Chile, Casilla 306, Santiago 22, Chile}
\affiliation{Chinese Academy of Sciences South America Center for Astronomy, National Astronomical Observatories, CAS, Beijing 100012, China}

\author{Jianfeng Wu}
\affiliation{Department of Astronomy, Xiamen University, Xiamen, Fujian 361005, China}

\begin{abstract}

We report the discovery of an extreme X-ray flux rise (by a factor of $\gtrsim$ 20) of the weak-line quasar SDSS J153913.47+395423.4 (hereafter SDSS J1539+3954) at $z =$ 1.935. SDSS J1539+3954 is the most-luminous object among radio-quiet type~1 AGNs where such dramatic X-ray variability has been observed. Before the X-ray flux rise, SDSS J1539+3954 appeared X-ray weak compared with the expectation from its UV flux; after the rise, the ratio of its X-ray flux and UV flux is consistent with the majority of the AGN population. We also present a contemporaneous HET spectrum of SDSS J1539+3954, which demonstrates that its UV continuum level remains generally unchanged despite the dramatic increase in the X-ray flux, and its \civ\ emission line remains weak. The dramatic change only observed in the X-ray flux is consistent with a shielding model, where a thick inner accretion disk can block our line of sight to the central X-ray source. This thick inner accretion disk can also block the nuclear ionizing photons from reaching the high-ionization broad emission-line region, so that weak high-ionization emission lines are observed. Under this scenario, the extreme X-ray variability event may be caused by slight variations in the thickness of the disk. This event might also be explained by gravitational light-bending effects in a reflection model.

\end{abstract}


\section{Introduction} \label{sec:intro}

Weak-line quasars (WLQs; e.g., \citealt{Fan1999,DS2009,Plotkin2010}) are a notable group of active galactic nuclei (AGNs) largely discovered by the Sloan Digital Sky Survey (SDSS; \citealt{York2000}). While typical quasars show strong and broad line emission in the optical/ultraviolet (UV), WLQs are type~1, radio-quiet quasars with weak or no emission lines.
WLQs with \civ\ rest-frame equivalent widths (REWs) $\lesssim 10$ \AA\ deviate negatively at $\gtrsim 3\sigma$ levels from the mean \civ\ REW of SDSS quasars; at the same time, there is no such population that deviates positively at $\gtrsim 3\sigma$ levels from the mean of the \civ\ REW distribution \citep[e.g.,][]{DS2009,Wu2012}.

WLQs have also exhibited remarkable X-ray properties.
For typical radio-quiet quasars without broad absorption lines (BALs), the \hbox{X-ray-to-optical} power-law slope parameter (\aox)\footnote{\aox\ is the power-law slope connecting the monochromatic luminosities at rest-frame 2500 \AA\ and 2 keV; \hbox{$\alpha_{\rm ox} = 0.384 \log(L_{\rm 2~keV}/L_{2500~\mathring{\rm{A}}})$}.} follows a correlation with 2500 \AA\ monochromatic luminosity (\lopt; e.g., \citealt{Steffen2006,Just2007,Lusso2016}).
However, about half of the WLQs have notably lower \xray\ luminosities compared to the expectation from the \aox-\lopt\ relation \citep[e.g.,][]{Luo2015,Ni2018}.
For this half of the WLQ population, high apparent levels of intrinsic X-ray absorption, Compton reflection, and/or scattering have been suggested through X-ray stacking and spectral analyses \citep[e.g.,][]{Wu2011,Wu2012,Luo2015, Ni2018}.
Also, the notably steep \xray\ spectra of the other half of the WLQ population that is not X-ray weak indicate accretion at high Eddington ratios \citep[e.g.,][]{Luo2015,Marlar2018}.

However, WLQs have not generally been associated with extreme X-ray variability before.
For typical AGNs, the long-term variation of X-ray luminosity is generally within a factor of $\approx 2$, and seldom exceeds a factor of $\approx 10$ \citep[e.g.,][]{Gibson2012,Yang2016,Middei2017}. Also, the anti-correlation between the X-ray variability amplitude and the black-hole mass \citep[e.g.,][]{Ponti2012} suggests that extreme \hbox{X-ray} variability is not generally expected among luminous quasars that tend to host high-mass black holes \citep[e.g.,][]{Shen2012}.
Extreme X-ray variability events (\hbox{X-ray} luminosity variations by factors of $\gtrsim 10$) have been found among radio-loud AGNs that have jet-linked \hbox{X-ray} variability \citep[e.g.,][]{Carnerero2017}, BAL quasars that have absorption-linked X-ray variability \citep[e.g.,][]{Saez2012}, and changing-look AGNs that exhibit multiwavelength variability \citep[e.g.,][]{Oknyansky2019}. There are also extreme X-ray variability events among radio-quiet non-BAL AGNs that do not have simultaneous changes in the UV/optical.
These events have mostly been associated with narrow-line Seyfert 1 galaxies (NLS1s) that have small black-hole masses \citep[e.g.,][]{Liu2019}.

In this Letter, we report the discovery of an extreme X-ray variability event of a radio-quiet non-BAL WLQ, SDSS J153913.47+395423.4 (hereafter SDSS~J1539+3954) at $z = 1.935$. SDSS~J1539+3954 is a WLQ selected in \citet{Plotkin2010} for its weak \civ\ emission features.
It has a bolometric luminosity of \lbol\ $\approx 1.5 \times 10^{47}$ erg s$^{-1}$ \citep{Shen2011}, which makes it the most luminous radio-quiet AGN showing extreme X-ray variability.

\section{Observations and Data Analyses} \label{sec:data}
In Section~\ref{ssec:xray}, we present the two \chandra\ \xray\ observations obtained for SDSS J1539+3954 which exhibit strong variation in the \xray\ flux level; in Section~\ref{ssec:spec}, we present all the optical spectroscopic observations for SDSS~J1539+3954 which are generally consistent.

\subsection{\chandra\ X-ray Observation and Data Analyses} \label{ssec:xray}
We list in Table~\ref{tab:xray} the two X-ray observations of SDSS~J1539+3954.
These \chandra\ observations were performed with the Advanced CCD Imaging Spectrometer spectroscopic array (ACIS-S; \citealt{Garmire2003}) in VFAINT mode.
The new \chandra\ Cycle 21 observation is a part of our program to obtain deeper X-ray coverage for a set of WLQs (PI: W.~N.~Brandt).

\begin{figure}[h!]
\includegraphics[scale = 0.32]{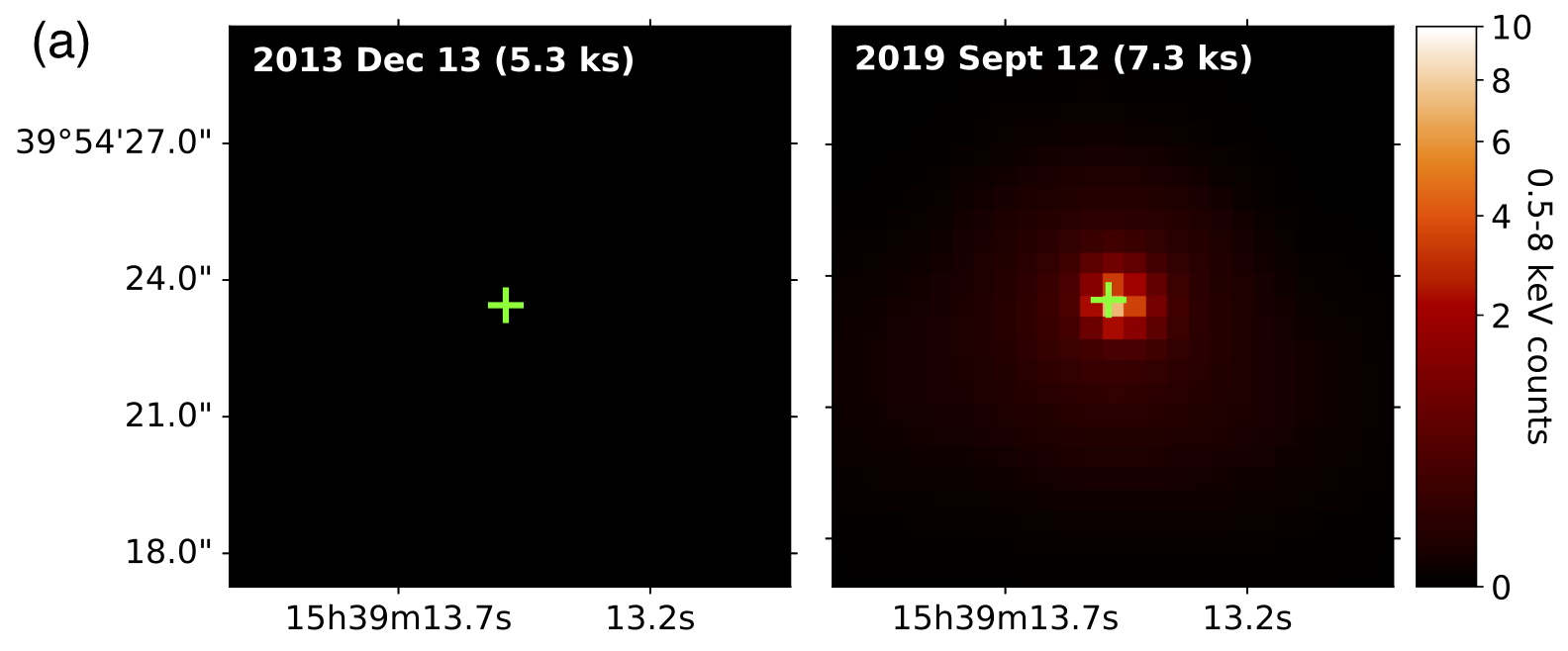}
\includegraphics[scale = 0.58]{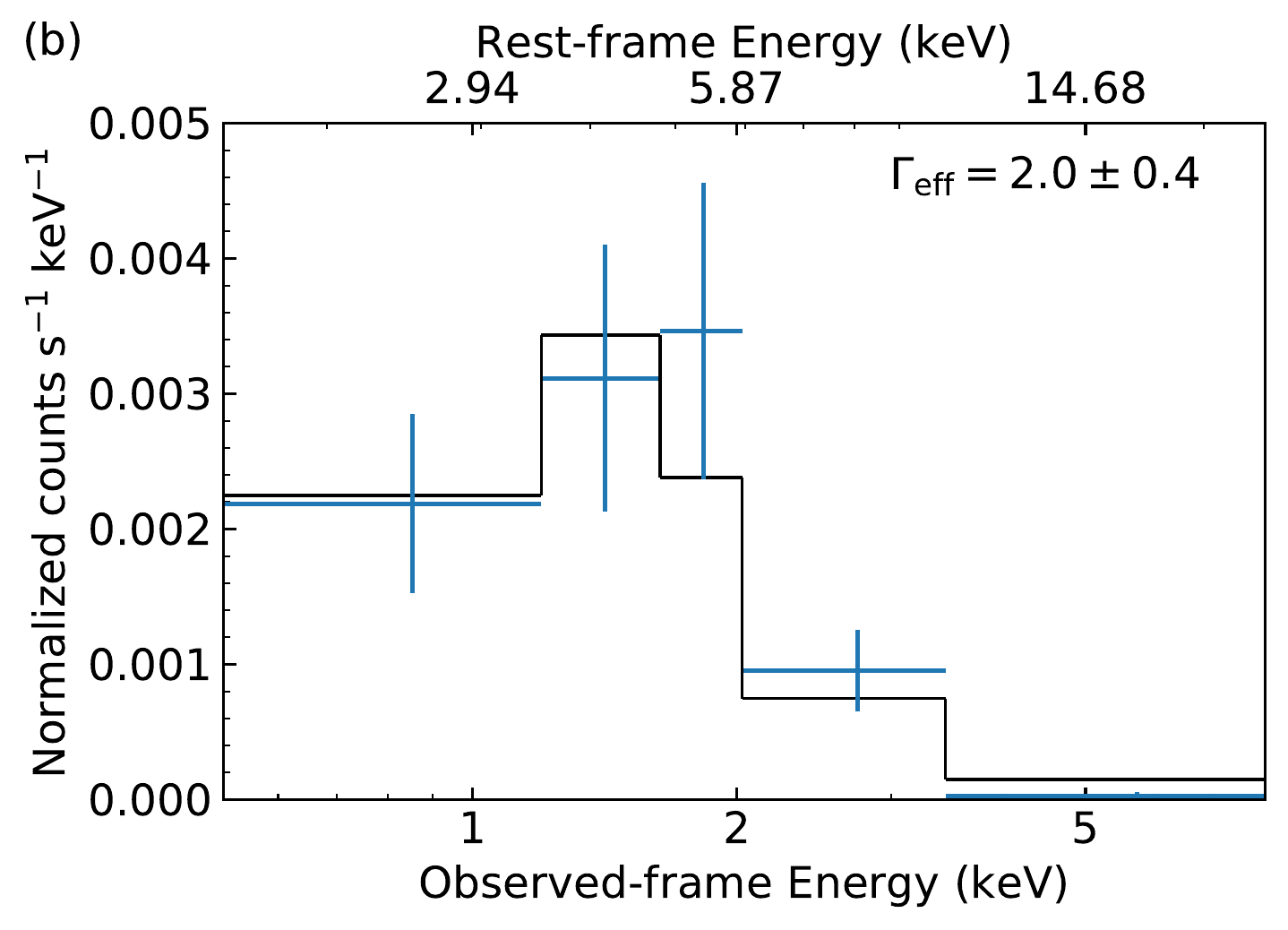}
\caption{(a) \chandra\ full-band (0.5--8 keV) images of SDSS~J1539+3954 in two different epochs (smoothed with CSMOOTH). SDSS~J1539+3954 was not detected in 2013 (see the left panel), and was detected with $\approx$ 44 counts in 2019 (see the right panel). (b) The full-band X-ray spectrum of SDSS~J1539+3954 from the 2019 \chandra\ observation (adjacent bins are combined until they provide a detection at a $> 3\sigma$ level for presentation purposes), shown with a folded \textit{phabs*powerlaw} model in XSPEC.}
\label{fig:xray}
\end{figure}

We processed the \chandra\ data using the Chandra Interactive Analysis of Observations (CIAO) tools \citep{Fruscione2006}, following the steps in \citet{Luo2015} and \citet{Ni2018}.
We first used the CHANDRA\_REPRO script and the DEFLARE script to create the cleaned event file, and then created images in the 0.5--2 keV (soft), 2--8 keV (hard), and 0.5--8 keV (full) bands.
Source positions are determined by WAVDETECT. In case of non-detection (see section~3 of \citealt{Ni2018} for the definition of non-detection), we adopt the SDSS position.
We performed aperture photometry in both the soft band and hard band: source counts were extracted from a 2$''$ radius circular aperture centered on the source position, and background counts were extracted from a source-free annular region with a 10$''$ inner radius and a 40$''$ outer radius. 
When the source is undetected, the upper limits on the source counts were derived using the 90\% confidence-level table in \citet{Kraft1991}.
As we can see in Figure~\ref{fig:xray}a, SDSS~J1539+3954 was not detected in the first 5.3~ks \chandra\ observation, and was detected with $\approx 44$ counts in total in the second 7.3 ks \chandra\ observation.

\begin{deluxetable*}{cccccccccc}
\tablecaption{X-ray observations and derived properties of SDSS~J1539+3954\label{tab:xray}}
\tablewidth{0pt}
\tablehead{
\colhead{Observation} & \colhead{Observation} & \colhead{Exposure} & \colhead{Soft-band} & \colhead{Hard-band}  
& \colhead{Band}  & \colhead{$\Gamma_{\rm eff}$} & \colhead{$f_{\rm 2~keV}$} & \colhead{\aox} & \colhead{\daox($\sigma$)} \\ 
\colhead{ID}                 & \colhead{Start Date}    & \colhead{Time} & \colhead{Counts}     & \colhead{Counts}        
& \colhead{Ratio}  & \colhead{}                                  & \colhead{}& \colhead{}& \colhead{} 
}
\decimalcolnumbers
\startdata
14948 & 2013 Dec 13 & 5.3 & $< 2.4$ & $< 2.5$ & - & - & $<0.83$ & $<-2.18$ & $< -0.46~(3.17)$ \\
22528 & 2019 Sept 12 & 7.3 & $29.1^{+6.6}_{-5.5}$ & $15.2^{+5.3}_{-4.1}$ & $0.52^{+0.21}_{-0.14}$ & $2.0^{+0.4}_{-0.4}$ & 16.58 &  $-1.68$ &$0.04~(0.26)$ \vspace{0.2 mm}
\enddata
\tablecomments{ (1) \chandra\ observation ID. (2) Observation start date. (3) Background-flare cleaned effective exposure time in the 0.5--8 keV band in units of ks. (4)/(5) Aperture-corrected source counts in the soft (0.5--2 keV)/hard (2--8 keV) band. An upper limit at a 90\% confidence level is given if the source is not detected. (6) Ratio between the soft-band and hard-band counts. ``-" indicates that the source is undetected in both bands. (7) 0.5--8 keV effective power-law photon index. ``-" indicates that $\Gamma_{\rm eff}$ cannot be constrained. (8) Rest-frame 2 keV flux density in units of 10$^{-32}$ erg cm$^{-2}$ s$^{-1}$ Hz$^{-1}$. (9) Measured \aox\ values. (10) Difference between the measured \aox\ and the expected \aox\ from the Just et al. (2007) \aox-\lopt\ relation. The statistical significance of this difference measured following table~5 of \citet{Steffen2006} is given in the parentheses.}
\end{deluxetable*}

\subsection{Spectroscopic Observations and Data Analyses} \label{ssec:spec}
We list in Table~\ref{tab:uv} all the optical spectroscopic observations of SDSS J1539+3954.
SDSS J1539+3954 was observed by SDSS-I/II on 2004 June 15, and the Baryon Oscillation Spectroscopic Survey of SDSS-III (BOSS; \citealt{Dawson2013}) on 2012 Apr 29.
The reduced SDSS and BOSS spectra were downloaded directly from the SDSS data archive.

\begin{figure*}[]
\plotone{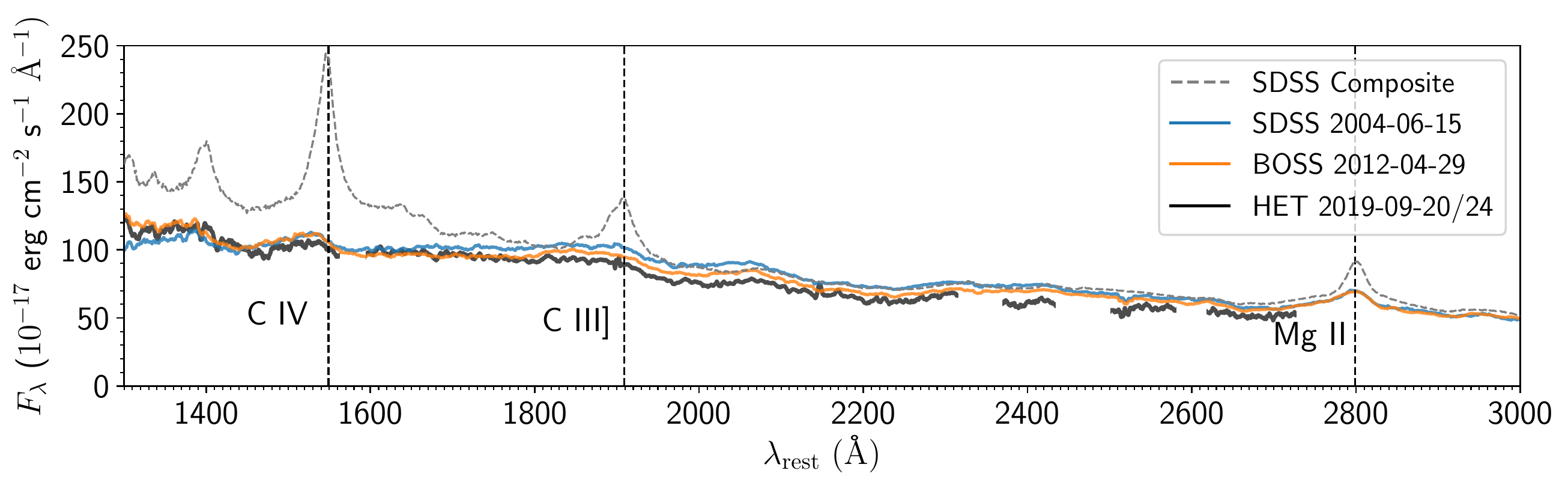}
\caption{HET spectrum of SDSS J1539+3954 taken $\approx 2.7$/4.1 rest-frame days after the discovery of the extreme X-ray variability event (displayed together with the earlier SDSS spectra). We mask the areas where the HET spectrum suffers from channel discontinuities and telluric absorption (G. Zeimann 2019, private communication). In general, the UV continuum and emission-line properties do not exhibit significant changes as in the X-ray. The SDSS quasar composite spectrum from \citet{VB2001} is scaled to the 2004 SDSS spectrum of SDSS J1539+3954 at rest-frame 2240 \AA\ and plotted in the background for comparison.}
\label{fig:spec}
\end{figure*}

After SDSS J1539+3954 was found to exhibit large \hbox{X-ray} variability, we promptly observed this object again with the Low-Resolution Spectrograph-2 (LRS2; \citealt{Chonis2014}) on the Hobby-Eberly Telescope (HET; \citealt{Ramsey1998}).
The observation with the blue arm of LRS2 (LRS2-B) was performed 8 days ($\approx$~2.7 days in the rest frame) after the \chandra\ observation, and the observation with the red arm of LRS2 (LRS2-R) was performed 12 days  ($\approx$ 4.1 days in the rest frame) after the \chandra\ observation. 
The LRS2 spectra were reduced with the HET pipeline \texttt{panacea}\footnote{https://github.com/grzeimann/Panacea}.
The current version of the HET pipeline cannot correct for telluric absorption, and there are still some channel discontinuities. The flux calibration is estimated to have an uncertainty around $20\%$ (G. Zeimann 2019, private communication). However, it is still sufficient for us to probe the basic properties of the UV continuum and emission lines.
In Figure~\ref{fig:spec}, we display all the spectra obtained for SDSS J1539+3954 together. They do not show noticeable variations.

\begin{deluxetable*}{cccccccc}
\tablecaption{Spectroscopic observations of SDSS J1539+3954 and the measured UV emission-line properties \label{tab:uv}}
\tablewidth{0pt}
\tablehead{
\colhead{Date} & Instrument & Spectral Coverage     & Exposure Time & S/N &  \colhead{\civ\ REW} & \colhead{\civ\ Blueshift} & \colhead{\civ\ FWHM}  \\
\colhead{}        &                    &  (\AA)          &   (s)                                    &        &(\AA)                            & \colhead{(km s$^{-1}$)}  & \colhead{(km s$^{-1}$)}       
}
\decimalcolnumbers
\startdata
2004 June 15 & SDSS            & 3800--9200   &   2700  & 25.6 &  $7.6 \pm 1.3$  & $5230 \pm 670$  & $11570 \pm 790$ \\
2012 Apr 29 & BOSS            & 3650--10400 &   2700  & 30.4 &  $8.2 \pm 0.9$  & $5230 \pm 910$  & $12320 \pm 530$\\
2019 Sept 20 & HET/LRS2-B & 3700--7000   &   1500  &  30.0 & $4.9 \pm 1.5$  & $4900 \pm 400$  & $10210 \pm 540$ \\
2019 Sept 24 & HET/LRS2-R & 6500--10500 &   1140  &  12.3 & - & - & - \vspace{0.2 mm}
\enddata
\tablecomments{(1) Date of the spectroscopic observation. (2) Name of the instrument. (3) Observed-frame spectral coverage of the instrument. (4) Total exposure time in seconds. (5) The average S/N at \hbox{$\lambda_{\rm rest} = 1750$--1800 \AA} for SDSS, BOSS, and HET/LRS2-B spectra, and at $\lambda_{\rm rest} = $ 2650--2700 \AA\ for the HET/LRS2-R spectrum. (6)/(7)/(8) REW/blueshift/FWHM of the \civ\ $\lambda$1549 emission line. ``-'' indicates that the \civ\ emission line is not covered in the spectrum. }
\end{deluxetable*}

\vspace{-0.2cm}

\section{Results} \label{sec:results}
\subsection{X-ray Properties} \label{ssec-xprop}
For the new \chandra\ detection ($\approx 44$ counts) of SDSS~J1539+3954, we calculated the band ratio (the ratio of hard-band counts to soft-band counts) and its uncertainty with the code BEHR \citep{Park2006}. 
Assuming a power-law spectrum modified by Galactic absorption, we use the Portable, Interactive, Multi-Mission Simulator (PIMMS)\footnote{https://cxc.harvard.edu/toolkit/pimms.jsp} to derive the 0.5--8 keV effective power-law photon index ($\Gamma_{\rm eff}$) and its uncertainty from the band ratio. 
We have also performed a power-law fit with XSPEC \citep{Arnaud1996} using the Cash statistic \citep{Cash1979} and obtained consistent results (see Figure~\ref{fig:xray}b for the X-ray spectrum).
The results are listed in Table~\ref{tab:xray}. The $\Gamma_{\rm eff}$ value of SDSS J1539+3954 is consistent with that of typical luminous radio-quiet quasars ($\Gamma_{\rm eff} =1.8$--2.0; e.g., \citealt{Reeves1997,Shemmer2005,Just2007,Scott2011}).
We then derived the unabsorbed soft-band flux (which covers rest-frame 2 keV) with PIMMS from the soft-band net count rate and $\Gamma_{\rm eff}$, and calculated the rest-frame 2 keV flux density ($f_{2~{\rm keV}}$) from the flux and $\Gamma_{\rm eff}$. 
In the case of \hbox{X-ray} non-detection, we could set an upper limit on $f_{2~{\rm keV}}$ following the same method using the upper limit for the soft-band net count rate and $\Gamma_{\rm eff} = 2.0$ adopted from the case of X-ray detection. 
As can be seen in Table~1, $f_{2~{\rm keV}}$ varied by a factor of $\gtrsim 20$ between the two epochs.
We note that different reasonable assumptions for the $\Gamma_{\rm eff}$ value ($\boldmath \Gamma_{\rm eff} \approx 1.6$--2.4) in the case of X-ray non-detection will not substantially change the results, and the variation in $f_{2~{\rm keV}}$ is always extreme (by a factor of $\gtrsim 17$--24).

\subsection{UV Emission-Line and Continuum Properties} \label{ssec-uvprop}
In Figure~\ref{fig:spec}, we can see that the UV continuum level and emission-line profiles of SDSS J1539+3954 remain generally unchanged from 2004 to 2019, considering the flux uncertainty of the HET spectrum.
Our measurement of the redshift $z = 1.935 \pm 0.004$ based on the Mg~{\sc ii} line in the SDSS and BOSS spectra is consistent with the value reported in \citet{Luo2015}.
Here, the adopted redshift is the average of two measurement results (from the two spectra); the uncertainty is the difference between two measurements.
We measured the \civ\ REW, blueshift, and FWHM in the three different epochs, as \civ\ properties are found to be linked with X-ray weakness among typical quasars (e.g., \citealt{Gibson2008,Timlin2019}).
For each spectrum, we performed both the local and global continuum fits following \citet{Yi2019}. For the local continuum fit, we fit a power-law function to the two ``anchor'' regions located on the blue/red-wing ends of the \civ\ emission line; for the global continuum fit, we model the continuum with a reddened power-law function to fit spectral windows that are relatively free of emission/absorption features.
After subtracting the fitted continuum, we fit the \civ\ emission with two Gaussian components to measure its REW, blueshift, and FWHM.
The REW is calculated from the fitted profile normalized by the continuum; the blueshift is converted from the wavelength bisecting the cumulative total flux of the \civ\ emission; the FWHM is derived from the combination of the two fitted Gaussian components.  
The reported \civ\ REW, blueshift, and FWHM values are the mean values generated by the two different continuum-fitting methods, and the reported uncertainties are the combination of the uncertainty from Monte Carlo simulations and the difference between two methods. The measurement results are shown in Table~\ref{tab:uv}.
We can see that the \civ\ emission-line properties do not have significant variations in the three different epochs.
While the \xray\ flux of SDSS J1539+3954 has experienced a dramatic increase, it remains a WLQ.

\subsection{The Multiwavelength SED and X-ray-to-optical Properties}
\begin{figure}[]
\centering
\includegraphics[scale = 0.54]{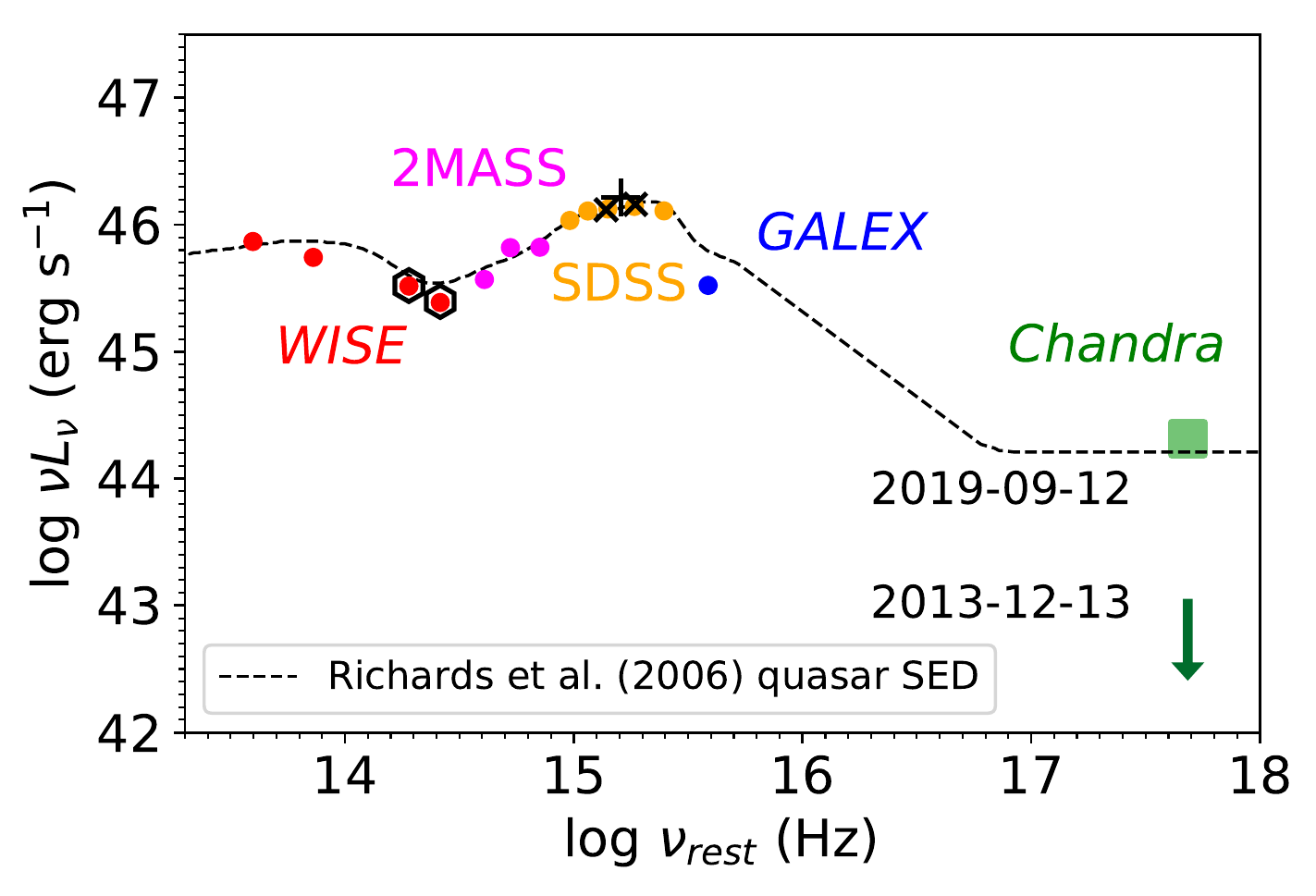}
\caption{SED of SDSS J1539+3954. The SED was scaled at rest-frame 3000 \AA\ to the composite SED (dashed line) of optically luminous quasars from \citet{Richards2006} by a factor of $\approx 0.3$. IR-to-UV SED data from \textit{WISE}, 2MASS, SDSS, and \textit{GALEX} are represented by red, magenta, orange, and blue points. The blue data point from \textit{GALEX} NUV falls below the composite SED due to Lyman-alpha absorption. The black crosses come from ZTF, the black plus comes from CRTS, and the black hexagons come from NEOWISE; the error bars showing the standard deviations of observed luminosities during the available monitoring are too small to be visible.
The green square shows the rest-frame 2~keV luminosity of SDSS J1539+3954 in 2019 from \chandra; the green arrow shows the rest-frame 2~keV luminosity upper limit (90\% confidence) of SDSS J1539+3954 in 2013 from \chandra.}
\label{fig:sed}
\end{figure}

In Figure~\ref{fig:sed}, we display the multiwavelength SED of SDSS J1539+3954, showing photometric data collected by the {\it Wide-field Infrared Survey Explorer} ({\it WISE}; \citealt{Wright2010}), Two Micron All Sky Survey (2MASS; \citealt{Skrutskie2006}), SDSS, {\it Galaxy Evolution Explorer} ({\it GALEX}; \citealt{Martin2005}), and \chandra, as well as the Catalina Real-time Transient Survey (CRTS; \citealt{Drake2009}), Zwicky Transient Facility (ZTF; \citealt{Bellm2019}), and Near-Earth Object Wide-field Infrared Survey Explorer project (NEOWISE; \citealt{Mainzer2014}).
CRTS monitored SDSS J1539+3954 from July 1, 2005 to Sept 28, 2013 (which is $\approx$ 2.5 months before the first \chandra\ observation of SDSS J1539+3954), and the results are reported in the $V$-band ($\approx$~1870~\AA\ in the rest frame); the Data Release 2 of ZTF includes the light curves of SDSS J1539+3954 from March 2018 to June 2019 in both the $g$-band and the $r$-band ($\approx$~1610~\AA\ and 2160~\AA\ in the rest frame);
the 2019 data release of NEOWISE contains the light curves of SDSS J1539+3954 from 2014 to 2018 at 3.4 $\mu m$ and 4.6 $\mu m$  ($\approx$~1.1~$\mu m$\ and 1.6~$\mu m$ in the rest frame).
No remarkable flux variation is detected by CRTS, ZTF, or NEOWISE, though we do observe modest variability in the ZTF and NEOWISE light curves (e.g., the ZTF light curves show variability of $\lesssim$ 0.1 mag).\footnote{The mean and standard deviation of the recorded CRTS $V$-band/ZTF $g$-band/ZTF $r$-band/NEOWISE 3.6~$\mu m$/NEOWISE 4.5~$\mu m$ magnitudes are $\approx 17.40 \pm 0.07$/$17.69 \pm 0.03$/$17.49 \pm 0.03$/$17.48 \pm 0.10$/$16.82 \pm 0.11$. These standard-deviation values are comparable to the mean measurement uncertainties in these five bands (0.09/0.03/0.02/0.09/0.11).}
The IR-to-UV SED of SDSS J1539+3954 is similar to those of typical quasars. 
The X-ray luminosity of SDSS J1539+3954 from the first \chandra\ observation is much lower than that of the composite SED of luminous quasars \citep{Richards2006}; the X-ray luminosity from the recent \chandra\ observation is roughly consistent with the composite SED.

We measured the \aox\ parameter for SDSS J1539+3954 in the two different X-ray epochs (see Footnote~1 for the equation).
%following the equation $\alpha_{\rm ox}=0.384 \log(L_{\rm 2~keV}/L_{2500~\mathring{\rm{A}}})$, where the ratio between \lx\ and \lopt\ is simply the ratio between \fx\ and \fopt.
%The ratio between monochromatic luminosities at rest-frame 2500 \AA\ and 2 keV is the ratio between rest-frame 2500 \AA~and 2 keV flux densities, \fopt\ and \fx.
As discussed before, the UV luminosity of SDSS~J1539+3954 does not exhibit significant variability during the long-term monitoring, and the spectroscopic observations of SDSS~J1539+3954 in three different epochs show roughly consistent UV continuum level. 
Thus, when calculating \aox, we adopt the \fopt\ value measured from the SDSS spectrum \citep{Shen2011} for the two different X-ray epochs consistently.
With the \fx\ value/upper limit obtained in Section~\ref{ssec-xprop}, we can calculate the \aox\ value/upper limit for the two epochs.
From the empirical \hbox{\aox--$L_{\rm 2500~\AA}$} relation for typical quasars \citep[e.g.,][]{Just2007}, we derived an expected value of \aox\ for SDSS J1539+3954.
Then, we measured the difference between the observed \aox\ and the expected \aox\ as \daox. 
The \aox\ and \daox\ results are listed in Table~\ref{tab:xray}.
\daox\ provides a measurement of \hbox{X-ray} weakness compared with the expected X-ray flux, as the ratio between the observed X-ray flux and expected \hbox{X-ray} flux is simply $10^{\Delta\alpha_{\rm OX}/0.384}$.
The first \chandra\ observation of SDSS J1539+3954 demonstrates that it is X-ray weak by a factor of $\approx 16$, and the second \chandra\ observation shows that SDSS J1539+3954 now follows the empirical \aox-\lopt\ relation well (lying $\approx 0.3\sigma$ above the relation).\footnote{Alternatively, if we compare the X-ray flux level with the expectation from the X-ray to MIR relation in \citet{Stern2015} (here we obtain the MIR luminosity from interpolating the {\it WISE} data shown in Figure~\ref{fig:sed}), the first \chandra\ observation is also X-ray weak by a factor of $\approx 16$, and the second \chandra\ observation follows the empirical relation well.}

\section{Discussion} \label{sec:discuss}

\subsection{The Second Discovery of Extreme X-ray Variability among WLQs}
The extreme X-ray variability discovered from SDSS~J1539+3954 is the second example of extreme \hbox{X-ray} variability among WLQs.
PHL 1092 is a $z = 0.40$ radio-quiet quasar with log $L_{\rm bol} \approx$ 46.65, and it was the first WLQ found to exhibit extreme \hbox{X-ray} variability (by a factor of $\approx 260$; e.g., \citealt{Miniutti2012}).
As summarized in \citet{Liu2019}, at high luminosities ($L_{\rm bol}$ $>$ $10^{46}$ erg s$^{-1}$), PHL 1092 was the only radio-quiet non-BAL quasar known to show such extreme X-ray variability, and like SDSS~J1539+3954 it varied between an X-ray weak state and an X-ray normal state.
SDSS~J1539+3954 has log $L_{\rm bol} \approx$ 47.17 \citep{Shen2011}, which is even more luminous (by a factor of $\approx 3$) than PHL 1092.
The discovery of extreme \hbox{X-ray} variability from SDSS J1539+3954, combined with the PHL~1092 results, suggests that weak UV emission lines may be a good indicator for finding extreme X-ray variability events among luminous radio-quiet quasars.

\subsection{Possible Explanations for Extreme X-ray Variability Events} \label{ssec-ps}

We have proposed a thick inner accretion-disk model to explain the multiwavelength properties of WLQs (e.g., figure~1 of \citealt{Ni2018}), which also has the potential to explain this extreme X-ray variability event. Simulations and analytical models suggest that for quasars with high Eddington ratios, geometrically thick inner accretion disks with high column densities are expected \citep[e.g.,][]{Abramowicz1988,Jiang2014,Jiang2019,Wang2014}.
Dense outflows arise with these thick disks in simulations \citep[e.g.,][]{Jiang2014,Jiang2019}, and we implicitly include any associated dense outflow within the term ``thick disk" throughout.
The thick inner accretion disk can prevent ionizing X-ray/EUV photons from reaching an equatorially concentrated high-ionization broad emission-line region (BLR), while the UV/optical photons from the disk itself remain unobscured. Thus, weak high-ionization emission lines are observed.
When our line of sight (to the central X-ray source) intercepts the thick inner accretion disk for a given WLQ, we observe an \hbox{X-ray} weak WLQ; when it misses this shield, we observe an \hbox{X-ray} normal WLQ.

As predicted in section~6.2 of \citet{Luo2015}, in the context of this model, the observed extreme X-ray variability event could be caused by a slight change in the thickness of the disk that moved across our line of sight.\footnote{\citet{Liu2019} proposed that a change in the height/size of the corona can also cause extreme X-ray variability in the presence of the thick disk via the same mechanism (as changes in both the disk thickness and the corona height/size are changes in the relative positions of the disk edge and the central X-ray source), though the detailed physics of the corona and how the corona varies its height/size is not generally understood.}
This could arise due to rotation of a thick inner disk that is somewhat azimuthally asymmetric, or alternatively due to small changes in the inner-disk structure itself (e.g., see figure~3 of \citealt{Jiang2019}).
If the change in the disk thickness is slight, there will be no significant change in the BLR illumination and UV/optical photon emission to affect the UV emission-line and continuum properties.\footnote{We note that it is also possible that the observed weak \civ\ line emission in the HET spectrum does not fully correspond to the \hbox{X-ray} normal state, as the average time lag between the \civ\ line emission and the ionizing continuum could reach $\approx 1$--2 years in the observed frame for SDSS~J1539+3954 \citep[e.g.,][]{Grier2019}. We think this possibility is not very likely, as the X-ray flux could have risen at any time in the six-year gap between the two \chandra\ observations, and while the closer half of the BLR responds to ionizing photons more promptly than the expected average lag, we do not observe any sign of stronger \civ. Future spectroscopic monitoring is needed to give a definite answer.}

\citet{Ross2005} have also proposed a reflection model that has the potential to explain this extreme X-ray variability event, where the \hbox{X-ray} variability is driven by gravitational light-bending effects. Such effects can change the amount of X-ray emission reaching an observer when the distance between the primary \xray\ source and the central black hole changes, causing the observed X-ray variability \citep{Miniutti2004}. However, in the context of this reflection model, it is not clear why extreme X-ray variability would have any particular link with WLQs.

%\vspace{-0.2cm}
\section{Conclusions and Future Work} 
We have reported a dramatic increase in the \xray\ flux of a luminous WLQ, SDSS J1539+3954, by a factor of $\gtrsim 20$.
We obtained a contemporaneous HET spectrum of SDSS~J1539+3954 after observing this X-ray flux rise with \chandra.
We found that its overall UV continuum and emission-line properties do not show significant changes compared with previous SDSS and BOSS spectra. 

We propose that the extreme X-ray variability event can be explained by a thick inner accretion-disk model with a slight change in the thickness of the disk, or a reflection model where strong gravitational light bending occurs.
We also note that the most luminous radio-quiet AGNs showing extreme X-ray variability are all WLQs, which may suggest a link between weak emission lines and extreme X-ray variability.

X-ray monitoring of a sample of WLQs will help constrain the frequency, duration, and amplitude of such extreme X-ray variability events, which can probe the possible scenarios listed in this paper
and thus clarify AGN accretion physics in general. For example, if extreme X-ray variability is caused by slight changes in the disk thickness, then this type of event should be relatively rare as we only expect such events to occur at viewing angles close to skimming the top of the thick inner disk. The smaller the change, the lower the frequency of extreme X-ray variability events we will observe.

%\clearpage
\acknowledgments
We thank HET astronomers Steven Janowiecki and Sergey Rostopchin for help obtaining the HET spectra.
We thank Greg Zeimman for help reducing the HET spectra.
QN, WNB, WY, and JDT acknowledge support from CXC grant GO8-19076, NASA ADP grant 80NSSC18K0878, and NSF grant AST-1516784.
BL acknowledges NSFC grant 11673010.
FV acknowledges financial support from CONICYT and CASSACA through the Fourth call for tenders of the CAS-CONICYT Fund, CONICYT grants BasalCATA AFB-170002.
PH acknowledges NSERC grant 2017-05983.

\bibliography{ex_wlq_var}
\bibliographystyle{aasjournal}

\end{document}